\documentclass{article}

\usepackage{PRIMEarxiv}

\usepackage[utf8]{inputenc} 
\usepackage[T1]{fontenc}    
\usepackage{hyperref}       
\usepackage{url}            
\usepackage{booktabs}       
\usepackage{amsfonts}       
\usepackage{nicefrac}       
\usepackage{microtype}      
\usepackage{lipsum}
\usepackage{fancyhdr}       
\usepackage{graphicx}       
\graphicspath{{media/}}     

\title{CPAISD: Core-Penumbra Acute Ischemic Stroke Dataset
}

\author{
  D. Umerenkov \\
  Sber AI Lab \\
  \texttt{d.umerenkov@gmail.com} \\
  \and
   \textbf{S. Kudin} \\
  Sber AI Lab \\
  \texttt{kudin.stepan@yandex.ru} \\
  \and
   \textbf{M. Peksheva} \\
  City hospital 40 of the Saint Petersburg Resort district  \\
  \texttt{pekshevam@gmail.com} \\
  \and
   \textbf{D. Pavlov} \\
  City hospital 40 of the Saint Petersburg Resort district \\
  \texttt{dg.pavlov.ctmri@gmail.com} \\
}

\begin{document}
\maketitle

\begin{abstract}
We introduce the CPAISD: Core-Penumbra Acute Ischemic Stroke Dataset, aimed at enhancing the early detection and segmentation of ischemic stroke using Non-Contrast Computed Tomography (NCCT) scans. Addressing the challenges in diagnosing acute ischemic stroke during its early stages due to often non-revealing native CT findings, the dataset provides a collection of segmented NCCT images. These include annotations of ischemic core and penumbra regions, critical for developing machine learning models for rapid stroke identification and assessment. By offering a carefully collected and annotated dataset, we aim to facilitate the development of advanced diagnostic tools, contributing to improved patient care and outcomes in stroke management. Our dataset's uniqueness lies in its focus on the acute phase of ischemic stroke, with non-informative native CT scans, and includes a baseline model to demonstrate the dataset's application, encouraging further research and innovation in the field of medical imaging and stroke diagnosis.
\end{abstract}


\section{Introduction}

Stroke, a leading cause of long-term disability and the second leading cause of death globally \cite{o2016global}, presents a significant challenge in medical imaging and diagnosis. The ability to rapidly and accurately diagnose stroke and determine the affected volumes is paramount in selecting appropriate treatment strategies to mitigate the devastating consequences of this condition. The traditional approach employs a multi-stage imaging protocol, beginning with a Native (Non-Contrast) CT scan of the brain, followed by more specialized scans such as CT Angiography (CTA) of the Brachiocephalic Arteries and CT Perfusion (CTP) Imaging of the brain \cite{wannamaker2019multimodal} . Additionally, Magnetic Resonance Imaging (MRI) is a reliable diagnostic tool for stroke. MRI offers detailed brain imaging, aiding in precise stroke identification and assessment. However, its availability is typically limited to large hospitals, making it less accessible in many regions. The multi-tiered process of CT or MRI, while comprehensive, is time-consuming and not universally accessible, posing significant limitations in the acute stroke setting where rapid intervention is vital.

The initial phase of an ischemic stroke is crucial for diagnosis, yet it frequently poses significant challenges. Native CT scans, the first line of imaging, often fail to reveal early indicators of ischemia, such as the blurring of the boundary between grey and white matter or the flattening of brain sulci. This absence of detectable changes can significantly delay the diagnosis and initiation of treatment, negatively impacting the recovery prospects of patients. While later imaging stages like CT Angiography (CTA) and CT Perfusion (CTP) are critical for assessing the presence of major arterial blockages and the extent of brain tissue damage, they require more time to perform and may not be readily accessible, particularly in resource-limited or rural healthcare settings.

Imaging models capable of reliably segmenting stroke areas using only Non-Contrast CT scans have significant implications for the management of stroke patients. Most importantly they enable faster diagnosis and treatment, crucial in time-sensitive situations. Their use would expand accessibility to stroke diagnostics, particularly in areas where advanced imaging technology is not readily available. Such models would also reduce patients' exposure to contrast agents and radiation, minimizing potential side effects. In terms of hospital management, they would enhance resource utilization, allowing for more efficient use of medical imaging equipment. Lastly, these models would aid in better triage and transfer decisions, ensuring patients receive the most appropriate care quickly.

The availability of open datasets containing segmented images of acute ischemic stroke is crucial for the development and validation of stroke detection models using Non-Contrast CT scans. These datasets serve as a critical resource for researchers and developers, allowing them to train and refine algorithms capable of identifying and segmenting ischemic areas with high precision. The benefits of such datasets are manifold:
\begin{itemize}
    \item \textbf{Enhanced Model Training}: Open access to segmented NCCT images provides a diverse data pool for training machine learning models, ensuring they learn from a wide range of stroke presentations.
    \item \textbf{Improved Model Validation and Benchmarking}: These datasets allow for rigorous testing and comparison of different models, ensuring their reliability and accuracy in real-world clinical settings.
    \item \textbf{Promoting Equity in Healthcare}: Open datasets ensure that advancements in stroke detection are not limited to well-resourced institutions but are available for broader application, including in underserved regions.
    \item \textbf{Reducing Development Costs and Time}: Access to pre-segmented and curated datasets reduces the resources required for data collection and preparation, expediting the development of effective diagnostic tools.
\end{itemize}

In this context, we present a novel dataset comprising NCCT scans along with corresponding ischemic core and penumbra segmentation masks for patients with early ischemic stroke. An unique feature of this dataset is that for absolute majority of the cases the Native CT findings are non-revealing in the sense that a trained radiologist was not able to find ischemic stroke areas using NCCT data alone.

This dataset offers a unique opportunity to explore the potential of advanced imaging techniques in augmenting the diagnostic capabilities of Non-Contrast CT scans in early stroke stages. The inclusion of markup based on CTA and CT Perfusion data provides a rich source of information for developing and testing predictive models.

We also provide a baseline model developed to predict the findings of the more advanced CTA and CT Perfusion stages using only the data from the Non-Contrast CT scans. While this model has not been extensively optimised it demonstrates the capabilities of AI models to find ischemic stroke areas that are hard or impossible to find for human expert.

In summary, our dataset and baseline model represent a step forward in the quest to improve stroke care. By focusing on the challenging scenario of early ischemic stroke with non-informative NCCT scans, this work aims to provide crucial data for future research and development in stroke diagnostics, ultimately contributing to better patient care and outcomes in this critical area of medicine.

The dataset is available at \url{https://zenodo.org/records/10892316}

\section{Related work}
\subsection{Models}
Recent advancements in the field of medical imaging and machine learning have significantly contributed to the early identification and assessment of acute ischemic stroke using non-contrast computed tomography. In \cite{wu2019early}, researchers developed a radiomics-based patch classification model, achieving high identification accuracies in detecting ischemic strokes invisible to radiologists in ncCT. This method integrated spatial constraints to enhance identification performance, demonstrating the potential of radiomics models in clinical applications. Similarly, \cite{kuang2021eis} introduced EIS-Net, a multi-task learning approach for segmenting early infarct (EI) and scoring Alberta Stroke Program Early CT Score (ASPECTS) simultaneously. This network, using a 3D triplet convolutional neural network and a multi-region classification network, showed strong correlation with diffusion weighted MRI (DWI) in EI volume assessment and achieved a high intraclass correlation for ASPECTS scoring, indicating its precision in both segmentation and scoring tasks.

Further exploring this domain, \cite{el2022evaluating} evaluated the nnU-Net framework for segmenting early ischemic changes (EIC) in acute ishcemic stroke (AIS) patients. Despite the challenges, the model showed promising results, with Dice Similarity Coefficients and Intraclass Correlation Coefficients comparable to those of expert neuroradiologists, underscoring its potential as a decision-aid tool for clinicians. In a similar vein, \cite{cao2022deep} proposed an automated ASPECTS method using a neural network, which exhibited remarkable sensitivity and accuracy in scoring CT scans of AIS patients. This method not only correlated ASPECTS scores with patient prognosis but also indicated the size of the CTP core volume of an infarct. Lastly, \cite{kuang2019automated} aimed at automating ASPECTS scoring using a random forest classifier trained on texture features. This approach yielded high sensitivity, specificity, and area under the curve in individual ASPECTS region-level analysis, demonstrating its efficacy in objectively scoring NCCT images in AIS patients. 

Collectively, these studies demonstrate the evolving landscape of acute ischemic stroke diagnosis using non-contrast computed tomography, where machine learning and advanced imaging techniques are playing an increasingly pivotal role in enhancing accuracy and efficiency. For a thorough review on current trends in this area please refer to \cite{sheth2023machine}

\subsection{Datasets}
Several datasets have been developed to study acute ischemic stroke, encompassing both MRI and NCCT imaging techniques.

In the realm of MRI datasets, Isles 2015 \cite{maier2017isles} offers an essential benchmark for ischemic stroke lesion segmentation, emphasizing the precision in multispectral MRI analysis. Isles 2016 and 2017 \cite{winzeck2018isles} extend this work by focusing on predicting stroke lesion outcomes based on multispectral MRI data, contributing to a better understanding of patient prognoses. The ISLES2018 dataset \cite{hakim2021predicting} is particularly significant, featuring 156 CTP studies from acute ischemic stroke patients, with 64 designated for a hidden test set, presenting a unique challenge in predictive modeling. Further advancing the field, Isles 2022 \cite{hernandez2022isles} introduces a multi-center MRI dataset aimed at stroke lesion segmentation, highlighting technological advancements in imaging and data analysis. Complementing these, Sook-Lei Liew and colleagues have provided extensive contributions through datasets \cite{liew2018large} \cite{liew2022large}  that include a large, open-source collection of stroke anatomical brain images and manual lesion segmentations, thus broadening the scope for research and algorithm development in stroke imaging.

Compared to a number of MRI-focused datasets, there are only two NCCT datasets for acute ischemic stroke. The first, AISD \cite{liang2021symmetry}, comprises 397 NCCT scans of acute ischemic stroke, captured within 24 hours of symptom onset. These patients also underwent diffusion-weighted MRI within the same timeframe. The NCCT scans have a slice thickness of 5mm, with 345 used for training and validation, and 52 reserved for testing. Lesions are meticulously outlined on NCCT by medical professionals, using MRI as the reference standard. The second NCCT dataset, APIS \cite{gomez2023apis}, introduces a paired CT-MRI dataset meticulously built for ischemic stroke segmentation. It includes 96 studies from patients exhibiting stroke symptoms, divided into control (n=10) and ischemic stroke (n=86) groups. This dataset is notable for its methodical approach to data collection and stringent inclusion criteria.

\section{Dataset}
\subsection{Data acquisition}
In the Regional Vascular Center of State Budgetary Healthcare Institution №40 of St. Petersburg, 135 patients were selected during the period from 2017 to 2020. These patients underwent all three stages of CT examinations (native scanning, CT angiography, CT perfusion). All patients were admitted in the acute phase of ischemic stroke, within 24 hours of the onset of symptoms. The exact time from symptom onset was not generally known, a significant portion of the patients where either found unconscious by another people or woke up from sleep with stroke symptoms. There were 25 fatal outcomes among the selected patients.

23 patients were excluded from further processing due to reasons such as damaged DICOM data and motion artifacts. For subsequent processing, 112 patients were selected. Of these, 40 did not undergo subsequent native CT scanning on the first day, but had CT or MRI control on the second or third day; 41 patients were selected for endovascular treatment.

Out of the 112 selected patients:
\begin{itemize}
    \item 1 patient with MCA stroke at both hemispheres, contains ischemic core.
    \item 2 patients with ICA stroke at both hemispheres, contains ischemic core (all lethal).
    \item 4 patients with right ICA stroke, 3 patients contains ischemic core.
    \item 38 patients with right MCA stroke, 1 patient had lacunar stroke, 28  patients contains ischemic core.
    \item 1 patient with right ACA stroke, without ischemic core.
    \item 5 patients with left ICA stroke, all contains ischemic core.
    \item 54 patients with left MCA stroke, 2 patients has lacunar stroke, 37 patients patients contains ischemic core.
    \item 7 patients with stroke at basilar artery territory, 6 patients patients contains ischemic core.
\end{itemize}

Subsequently, a radiologist with 10 years of experience performed the markup of the native scanning series across all sections, comparing with CT perfusion maps, and delineated areas of penumbra and core (one within the other). The software used for processing was Horos.

\subsection{Data format}

The dataset contains 112 non-contrast cranial CT scans of patients with hyperacute stroke, featuring delineated zones of penumbra and core of the stroke on each slice where present. The data in the dataset are anonymized using the Kitware DicomAnonymizer \cite{dicom-anonymizer}, with standard anonymization settings, except for preserving the values of the following fields:
\begin{itemize}
    \item (0x0010, 0x0040) -- Patient's Sex
    \item (0x0010, 0x1010) -- Patient's Age
    \item (0x0008, 0x0070) -- Manufacturer
    \item (0x0008, 0x1090) -- Manufacturer’s Model Name
\end{itemize}

The patient's sex and age are retained for demographic analysis of the samples, and the equipment manufacturer and model are kept for dataset statistics and the potential for domain shift analysis.

The dataset is split into three folds:
\begin{itemize}
    \item Training fold (92 studies, 8,376 slices).
    \item Validation fold (10 studies, 980 slices).
    \item Testing fold (10 studies, 809 slices).
\end{itemize}

The dataset has the following structure:
\begin{itemize}
    \item \texttt{metadata.json} – dataset metadata
    \item \texttt{summary.csv} – metadata of each study in a CSV format table
    \item Part of the dataset (train, val, and test)
    \begin{itemize}
        \item Study
        \begin{itemize}
            \item \texttt{metadata.json} – study metadata in JSON format
            \item Slice
            \begin{itemize}
                \item \texttt{raw.dcm} – original slice file
                \item \texttt{image.npz} – slice in Numpy array format
                \item \texttt{mask.npz} – segmentation mask in Numpy array format
                \item \texttt{metadata.json} – slice metadata in JSON format
            \end{itemize}
        \end{itemize}
    \end{itemize}
\end{itemize}

The `metadata.json` at the root of the dataset has the following format:
\begin{itemize}
  \item \texttt{generation\_params} – dataset generation parameters:
  \begin{itemize}
    \item \texttt{test\_size} – proportion of the test part
    \item \texttt{val\_size} – proportion of the validation part
  \end{itemize}
  \item \texttt{stats} – statistical data:
  \begin{itemize}
    \item \texttt{common} – general statistical data:
    \begin{itemize}
      \item \texttt{train\_size\_in\_studies} – number of studies in the training part of the dataset.
      \item \texttt{train\_size\_in\_images} – number of slices in the training part of the dataset.
      \item \texttt{val\_size\_in\_studies} – number of studies in the validation part of the dataset.
      \item \texttt{val\_size\_in\_images} – number of slices in the validation part of the dataset.
      \item \texttt{test\_size\_in\_studies} – number of studies in the test part of the dataset.
      \item \texttt{test\_size\_in\_images} – number of slices in the test part of the dataset.
    \end{itemize}
    \item \texttt{train} – statistical data for the training part of the dataset:
    \begin{itemize}
      \item \texttt{min} – minimum pixel value.
      \item \texttt{max} – maximum pixel value.
      \item \texttt{mean} – average pixel value.
      \item \texttt{std} – standard deviation for all pixel values.
    \end{itemize}
  \end{itemize}
\end{itemize}

The `metadata.json` at the root of the study has the following format, if a field value is unknown, it is given as 'unknown':
\begin{itemize}
    \item \texttt{manufacturer} – manufacturer of the tomograph.
    \item \texttt{model} – model of the tomograph.
    \item \texttt{device} – full name of the tomograph (manufacturer + model).
    \item \texttt{age} – patient's age in years.
    \item \texttt{sex} – patient's sex. M – male, F – female.
    \item \texttt{dsa} – whether cerebral angiography was performed. true if yes, false if no.
    \item \texttt{nihss} – NIHSS score.
    \item \texttt{time} – time in hours from the onset of the stroke to the conduct of the study. Can be either a number or a range.
    \item \texttt{lethality} – whether the person died as a result of this stroke. true if yes, false if no.
\end{itemize}

The `summary.csv` contains the same fields as the `metadata.json` from the root of the study, plus two additional fields:
\begin{itemize}
    \item \texttt{name} – name of the study.
    \item \texttt{part} – part of the dataset in which the study is located.
\end{itemize}

\section{Baseline model}

We used the CPAISD dataset to train a baseline core and penumbra segmentation model. The baseline model is a segmentation network of the FPN \cite{lin2017feature} architecture with an efficientnet-b0 \cite{tan2019efficientnet} backbone, segmenting a single-channel input image into three classes. The network receives a single slice as input and outputs a three-channel mask, with class 0 representing the background, class 1 the stroke core, and class 2 the penumbra.

The network was trained using the Adam optimizer, with DiceLoss as the loss function, which only accounted for stroke core and penumbra. The learning rate scheduler used was ReduceLROnPlateau. The initial learning rate was 0.006 and batch size during training was 32. 
The input data was normalized using standard normalization - the mean value of the pixel across the training set is subtracted from each pixel of the image, and then it is divided by the standard deviation of the pixel values on the training set. For training augmentations we used horizontal flipping and random rotation within a range of -10 to 10 degrees.

The average value of the DICE 3D metric on the test part of the dataset is 0.2183.

Te model code and weights can be accessed at \url{https://github.com/sb-ai-lab/early_hyperacute_stroke_dataset} 

\begin{figure}
    \centering
    \includegraphics[width=\textwidth]{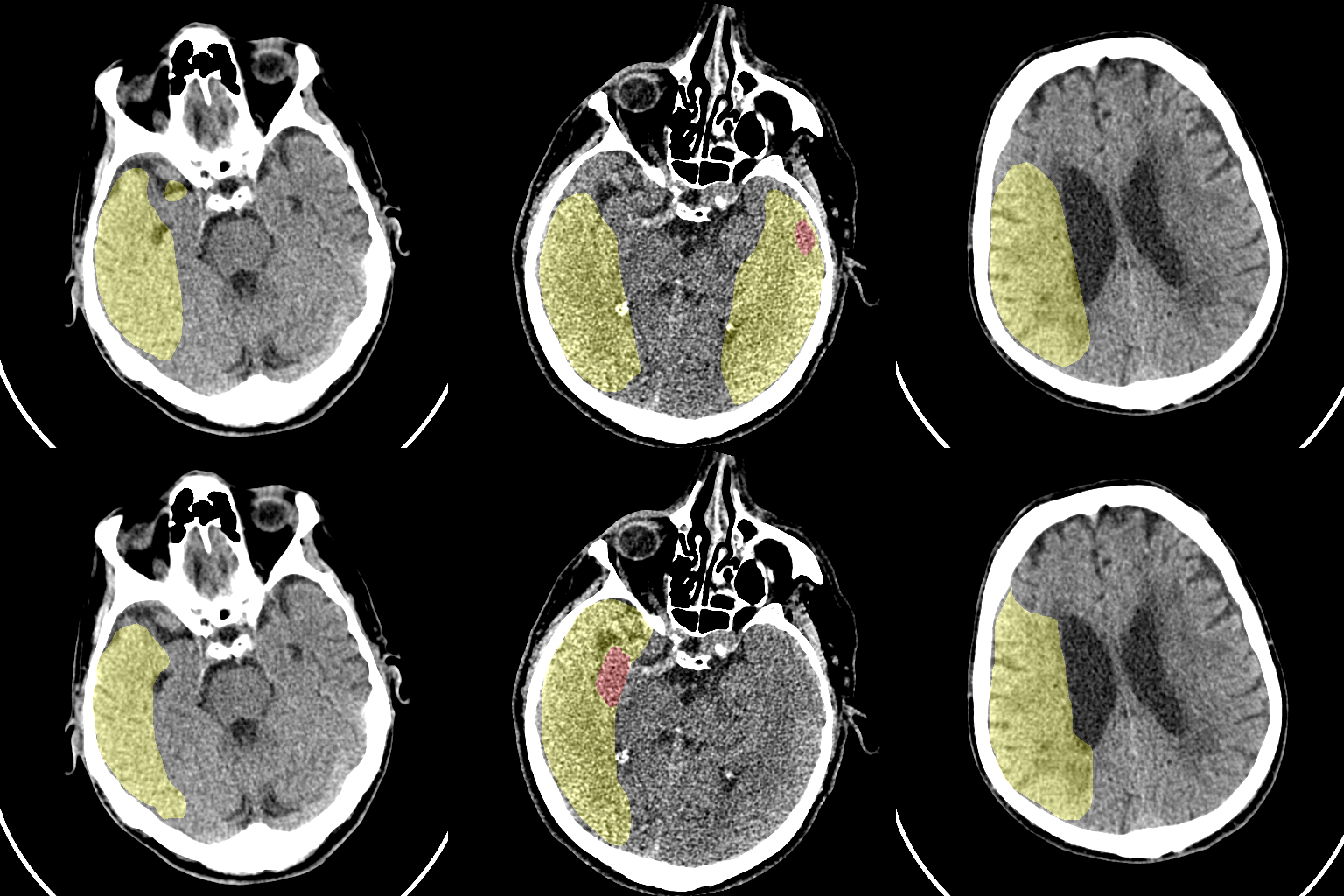}
    \caption{Baseline model prefictions. Predictions on the top row, ground truth in bottom. Core is shown in red, penumbra in yellow }
    \label{fig:example}
\end{figure}

The example of baseline model prediction and the ground truth is shown in Figure \ref{fig:example}. 

\section{Conclusion}

The advancement of medical imaging technologies, especially in the early detection and segmentation of ischemic stroke, hinges critically on the availability of comprehensive and detailed open access datasets. Our work introduces a novel segmentation dataset specifically designed for the important task of early ischemic stroke detection. This dataset enhances the breadth of resources available to researchers aiding in the development of advanced diagnostic tools for acute ischemic stroke. We hope that this dataset will serve as a foundational platform for future innovations in medical imaging and stroke diagnosis.

The uniqueness of our dataset is in its inclusion of segmentation annotations for both the core and penumbra regions of ischemic stroke in acute patients. This specificity provides an unprecedented level of detail, facilitating the development of more precise and sensitive diagnostic models. By capturing these critical regions within the acute phase of ischemic stroke, our dataset addresses a significant gap in the field, enabling researchers to explore new methodologies for early stroke detection and segmentation. 

We also provide a baseline model that demonstrates the practical application of our dataset. It is our hope that this model serves as a starting point for other researchers, encouraging the exploration of new algorithms and techniques in stroke detection and segmentation. By making our dataset and baseline model available to the wider research community, we aim to catalyze further advancements in this critical area of medical imaging. We expect that the availability of this specialized dataset will not only foster innovation but also significantly enhance the accuracy and efficiency of ischemic stroke diagnosis, ultimately improving patient care and outcomes in acute stroke management.


\bibliographystyle{unsrt}  
\bibliography{references}

\end{document}